\documentclass[12pt]{article}
\usepackage{graphics,color}
\begin{document}
\thispagestyle{empty}
\noindent\
\\
\\
\\
\begin{center}
\large \bf  Composite Weak Bosons and Dark Matter
\end{center}
\hfill
 \vspace*{1cm}
\noindent

\begin{center}
{\bf Harald Fritzsch}\\

Institute of Advanced Studies\\
Nanyang Technological University\\
Singapore\\

and

Department  f\"ur Physik\\ 
Ludwig-Maximilians-Universit\"at\\
M\"unchen, Germany \\

\vspace*{0.5cm}
\end{center}

\begin{abstract}

The weak bosons are bound states of two  
fermions and their antiparticles, denoted as haplons. The confinement scale 
of the associated gauge group SU(2) is of the order of 0.3 TeV.
Besides the weak bosons there exist also new bosons, an $SU(2)$-triplet and a 
singlet, with a mass of the order of 0.5 TeV. The neutral singlet boson is 
stable and provides the dark matter in the universe. 
 
\end{abstract}

\newpage

We assume that the weak bosons are composite particles. They consist 
of a lefthanded fermion and its antiparticle, 
which are denoted as "haplons".  A theory of this type was proposed in 1981 (see  ref.(1), also ref.(2,3,4,5,6)). 
The new confining chiral gauge theory is denoted as $QHD$. The haplons interact with each other through 
the exchange of massless gauge bosons.\\ 

Here we assume that the $QHD$ gauge group is the unitary group $SU(2)$. 
The simplest bound states of $QHD$ are bosons, either with 
haplon number H = 0, H = +2 or H = -2. The weak bosons are bound states with H = 0.\\  

The $QHD$ mass scale is given by a mass parameter $\Lambda_h$, 
which determines the size of the weak bosons. The  $QHD$ mass scale is about thousand times larger than the  $QCD$ 
mass scale (ref.(7,8)). The weak bosons are s-wave bound states of a haplon and an anti-haplon. The boson, observed recently 
at CERN (ref.(9,10)), is the lowest p-wave excitation of the neutral weak boson.\\ 

Two types of haplons are needed as constituents of the weak bosons, denoted by $\alpha$ and $\beta$.  
Their electric charges in units of e are:\\

\begin{equation}
h = \left( \begin{array}{l}
+\frac{1}{2}\\
-\frac{1}{2}\\
\end{array} \right) \ .
\end{equation}\\

The three weak bosons have the following internal structure:\\
\begin{eqnarray}
W^+ & = & \overline{\beta} \alpha \; , \nonumber \\
W^- & = & \overline{\alpha} \beta \; , \nonumber \\
W^3 & = & \frac{1}{\sqrt{2}} \left( \overline{\alpha} \alpha -
\overline{\beta} \beta \right) \; .
\end{eqnarray}\\

The neutral weak boson mixes with the photon. One obtains in first approximation the 
electroweak standard model (see ref. (7,8)).\\ 
\\

Now we consider the bound states with haplon number H=2. 
They form a triplet A and a singlet D of the electroweak symmetry:\\ 

\begin{eqnarray}
A^+ & = &( \alpha \alpha \;), \nonumber \\
A^3 & = &( \alpha \beta \;) , \nonumber \\
A^- & = &( \beta \beta \;) ,
\end{eqnarray}\\

\begin{eqnarray}
D & = &( \alpha \beta \;) .
\end{eqnarray}\\

We expect that the triplet bosons A have a higher mass than the neutral singlet boson D. The weak isospin is 
broken by the electromagnetic interaction. Thus the charged triplet bosons 
would have a slightly larger mass than the neutral triplet boson, due to the electromagnetic self energy.\\ 

The triplet bosons can decay into the 
neutral D-boson by emitting a virtual weak boson or a photon:\\
\\
$A^+ \Longrightarrow  (D + "W^+") $ ,\\
$A^- \Longrightarrow  (D + "W^-") $ ,\\
$A^3 \Longrightarrow  (D + "Z" ) $ ,\\
$A^3 \Longrightarrow  (D + \gamma) $ .\\

The D-boson is stable, since the haplon number is conserved, as the baryon number in $QCD$.\\ 
\\
The masses of these bosons are expected to be given by the confinement scale of the confining gauge group $SU(2)$, which is estimated to be of 
the order of 0.3 TeV (see ref. (7,8)). Thus the masses would be in the range between 0.3 TeV and 0.8 TeV.\\

Both the charged and neutral A-bosons as well as the neutral D-boson can be produced together with their antiparticles at the 
LHC-accelerator in CERN. The D-bosons cannot be observed directly, but a large missing energy should be observed.\\

The charged A-bosons would decay into a D-boson 
and a virtual charged weak boson, decaying into a lepton or quark pair. Thus one should observe in the LHC-experiments a large missing energy and 
a lepton pair or two quark jets. The neutral A-boson would decay into the D-boson by emitting a photon or a virtual Z-boson.\\

Thus far nothing has been observed. Probably the lower part of the range of the mass spectrum, mentioned above, is excluded by the LHC experiments. The masses of the three A-bosons and of the D-boson, if they exist, are probably in the range 0.5 TeV ... 0.8 TeV.\\ 

We interpret the stable D-boson as the particle, providing the dark matter in the universe. The properties of this  
particle are similar to a Weakly Interacting Massive Particle ("WIMP").\\ 

Shortly after the Big Bang the universe was filled with a dense gas of D bosons and anti-D bosons. 
The symmetry between D and anti-D bosons is expected to be slightly broken, as the symmetry between quarks and antiquarks. 
Besides the baryon-antibaryon asymmetry there would also be an asymmetry between D-bosons and anti-D bosons. Thus 
there would be slightly more D bosons than anti-D bosons (or vice-versa). The D bosons and the anti-D bosons annihilate, e.g. as follows:\\ 

$D + \overline{D} \Longrightarrow  (W^+ + W^-) $  
   
$D + \overline{D} \Longrightarrow  (W^+ + W^- + Z) $    

$D + \overline{D} \Longrightarrow  (W^+ + W^- + W^+ + W^-) $    

$D + \overline{D} \Longrightarrow  (Z + Z) $            
 
$D + \overline{D} \Longrightarrow  ( Z +  \gamma ) $      
 
$D + \overline{D} \Longrightarrow  (\gamma + \gamma) $.  \\

After the annihilation a gas of D-bosons is left in the universe, which provides the dark matter.\\ 

The average local density of the dark matter in our galaxy  
is estimated to (see ref. (11)):\\

 $\rho^{\rm local}_{\rm DM} = (0.39 \pm 0.03) ~{\rm GeV}{\rm cm}^{-3}$.\\ 
 
If the mass of the D bosons is 0.5 TeV, there would be about 780 D bosons in a cubic meter. For a higher mass 
the density would be less.\\ 

The D bosons are stable - the only way to observe them is the production 
of D bosons and anti-D bosons at the LHC.\\

\end{document}